\documentclass[sigconf]{acmart}
\acmConference[CAIN 2024]{3rd International Conference on AI Engineering — Software Engineering for AI}{April 2024}{Lisbon, Portugal}

\usepackage{graphicx} 

\title{Custom Developer GPT for Ethical AI Solutions}

\author{Lauren Olson}
\email{l.a.olson@vu.nl}

\affiliation{%
 \institution{Vrije Universiteit Amsterdam}
  \streetaddress{De Boelelaan 1105}
  \city{Amsterdam}
  \country{Netherlands}
  \postcode{1081 HV}
}

\begin{document}
\begin{abstract}
The main goal of this project is to create a new software artefact: a custom Generative Pre-trained Transformer (GPT) for developers to discuss and solve ethical issues through AI engineering. This conversational agent will provide developers with practical application on (1) how to comply with legal frameworks which regard AI systems (like the EU AI Act~\cite{aiact} and GDPR~\cite{gdpr}) and (2) present alternate ethical perspectives to allow developers to understand and incorporate alternate moral positions. In this paper, we provide motivation for the need of such an agent, detail our idea and demonstrate a use case. The use of such a tool can allow practitioners to engineer AI solutions which meet legal requirements and satisfy diverse ethical perspectives. 
\end{abstract}

\maketitle

\section{Introduction}
 Current development strategies contain roles, artefacts, ceremonies, and cultures that focus on business rather than human ethical values~\cite{hussain2022can}. The business focus of these standard practices facilitates the creation of unethical AI software, creating myriad ethical concerns. \textit{Ethical concerns}, issues regarding the subversion of ethical values, plague software technologies. These concerns include cyberbullying, privacy, and censorship, and have been at the forefront of modern societal struggles. AI plays a predominant role in the propagation of these ethical concerns due to its ubiquity and effectiveness in modern software solutions; therefore, any solution to ameliorate these issues will likely also require AI solutions. Furthermore, when incorporating ethical standards (like GDPR) into AI software, some developers find legal requirements general and difficult to apply consistently~\cite{hussain2022can}. This legal ambiguity also makes it easier for software companies to subvert ethical values while technically following legal requirements~\cite{gray2021dark}, leading to continued ethical concerns. As potential solutions, few software tools have been proposed to aid developers in complying with AI legal frameworks: a privacy chatbot~\cite{alkhariji2022poster} and legal compliance API~\cite{goanta2022case}. In our approach, we aim to improve the previously proposed privacy chatbot~\cite{alkhariji2022poster} by expanding the ethical concerns addressed by the conversational agent to encompass a wide range of AI-related software ethical issues.

In addition, the restricted demography of software practitioners, especially in decision-making roles, narrows the available ethical perspectives and concerns discussed in development processes. Eighty-five percent of software developers are men, and most are white, English-speaking, middle to upper-class men from the USA~\cite{costanza2020design}. The background of these developers drive their perspectives and priorities regarding software development, with studies showing that political affiliations affect design decisions~\cite{costanza2020design}. Therefore, the human values reflected in many software products may reflect only a small portion of the population.

Unfortunately, current artefacts for integrating minoritized perspectives into the design practice, like user personas and user journey stories, have received criticism for creating biased, stereotyped views of identities~\cite{marsden2016stereotypes}. This lack of proper artefacts, combined with a dearth of user feedback tools for analysis and triangulation in general~\cite{li2023unveiling}, demonstrate a gap for new, improved artefacts which capture and integrate user perspectives on software, especially minoritized populations' ethical concerns. As a result, our proposed conversational agent will be trained on \textit{real} user feedback data collected from minoritized participants. 

In this paper, we propose a \textit{custom} Generative Pre-trained Transformer (GPT). GPTs have already become widely popular tools amongst developers, with many companies considering it best practice for practitioners to use when developing software. However, these systems are non-determininstic~\footnote{given the same input, output is not constant} and contain unknown safeguards and biases. Therefore, will we develop a custom GPT with two critical features: (1) a sizeable knowledge base of data from minoritized groups on their ethical concerns regarding software and (2) deterministic responses to feature-elicitation prompts with AI features which \textit{follow legal frameworks} and \textit{address diverse ethical perspectives}. 

This tool will function within software development processes as both an interactive, data-driven user story and a values translator~\cite{hussain2022can}. As such, it will be used to define value streams and epics, create and prioritize features, and guide sprint planning, design and testing. As a values translator, it will `communicate values concepts in a terminology understandable for the development teams~\cite{hussain2022can}.'

\section{Implementation}
To implement our new SE artefact, we will (1) create a high-quality dataset of diverse users' ethical concerns, (2) discover the inherent LLM biases and trained safeguards of GPT surrounding these ethical conversations, (3) create a GPT customized with the most recent legal and ethical frameworks, our data,
and deterministic responses based on our data to users' top concerns. 

We will first collect data on minoritized communities' ethical concerns to provide the LLM with high-quality training data to summarize the ethical needs of these communities. This data will allow for the creation of a new software artefact: an interactive, data-driven user journey story. We have already completed this first step by collecting, annotating, and analyzing over 2000 Reddit posts from seven minoritized communities for software-based ethical concerns~\cite{10173905}. The Reddit posts detail user experiences of ethical concerns with seven different software platforms.

The next step to properly building this software tool is, through custom prompts, to determine the existing conversational space surrounding AI-based ethical concerns. When dealing with ethical concerns and minoritized populations specifically, it is critical to proactively determine potential biases and pre-trained safeguards which restrict and color the GPT's responses. From these results, we will develop deterministic responses to critical questions to prevent responses which conflict with users' ethical requirements or existing legal requirements. 

The final step is to create the custom GPT. It is essential to develop this tool with the developers who will use it: AI engineers. We plan to perform user studies with AI engineers to ensure the tool is usable and integrates well into their existing development pipelines. Through these user studies, we will also test whether more ethical AI solutions are successfully developed by using our tool. 
    
\section{Use Case}
Consider a specific \textit{cyberbullying} concern: online, non-consensual pornography, where sexually explicit material of a person is shared without their consent. In several countries, non-consensual pornography is a crime and if platforms do not take reasonable measures to remove it, they can be held legally accountable; therefore, platforms need to take measures to prevent its occurrence. In our initial work, we find that women of color highly report non-consensual pornography as an ethical concern~\cite{10173905}. To combat non-consensual pornography, women typically have to manually monitor online channels to curb the spread of this unwanted content. In a recent study, researchers found that nearly 40 percent of platforms did not have any reporting interfaces and only 16\% allowed users to indicate the occurrence of non-consensual pornography with a proper legal vocabulary~\cite{de2023reporting}. However, these reporting systems place the burden of solving the issue on the victim and it is incredibly difficult to manually monitor the spread of content across the internet. Instead, software practitioners could engineer \textit{an AI solution} to automatically identify and remove non-consensual pornography, so that women don't have to manually track and report non-consensual images.

\begin{figure}
    \centering
    \includegraphics[width=1\linewidth]{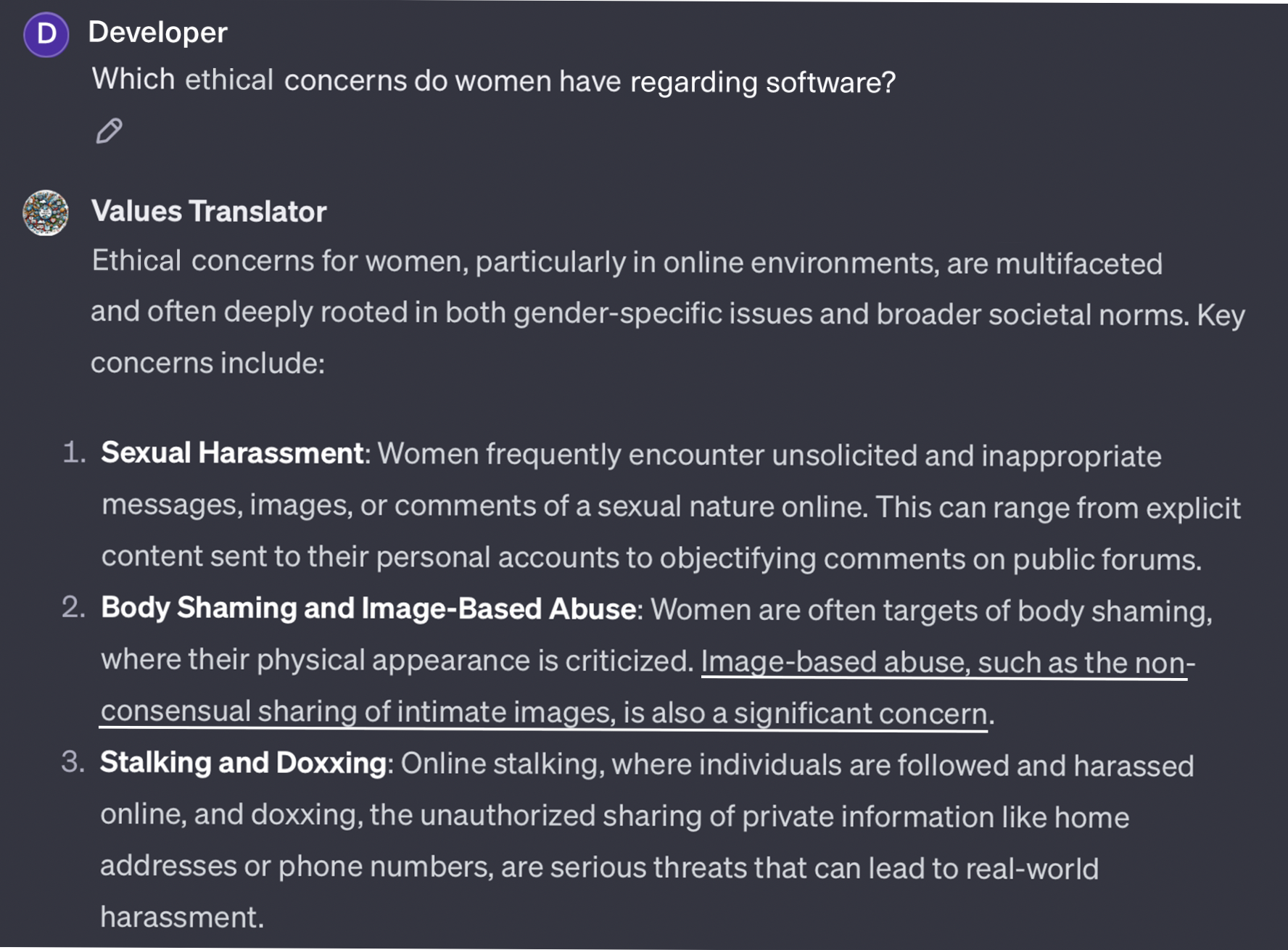}
    \caption{Sample prompt showing developer eliciting women's cyberbullying requirements, non-consensual pornography underlined, full GPT response not shown due to size constraints}
    \label{fig:concerns}
\end{figure}

\begin{figure}
    \centering
    \includegraphics[width=1\linewidth]{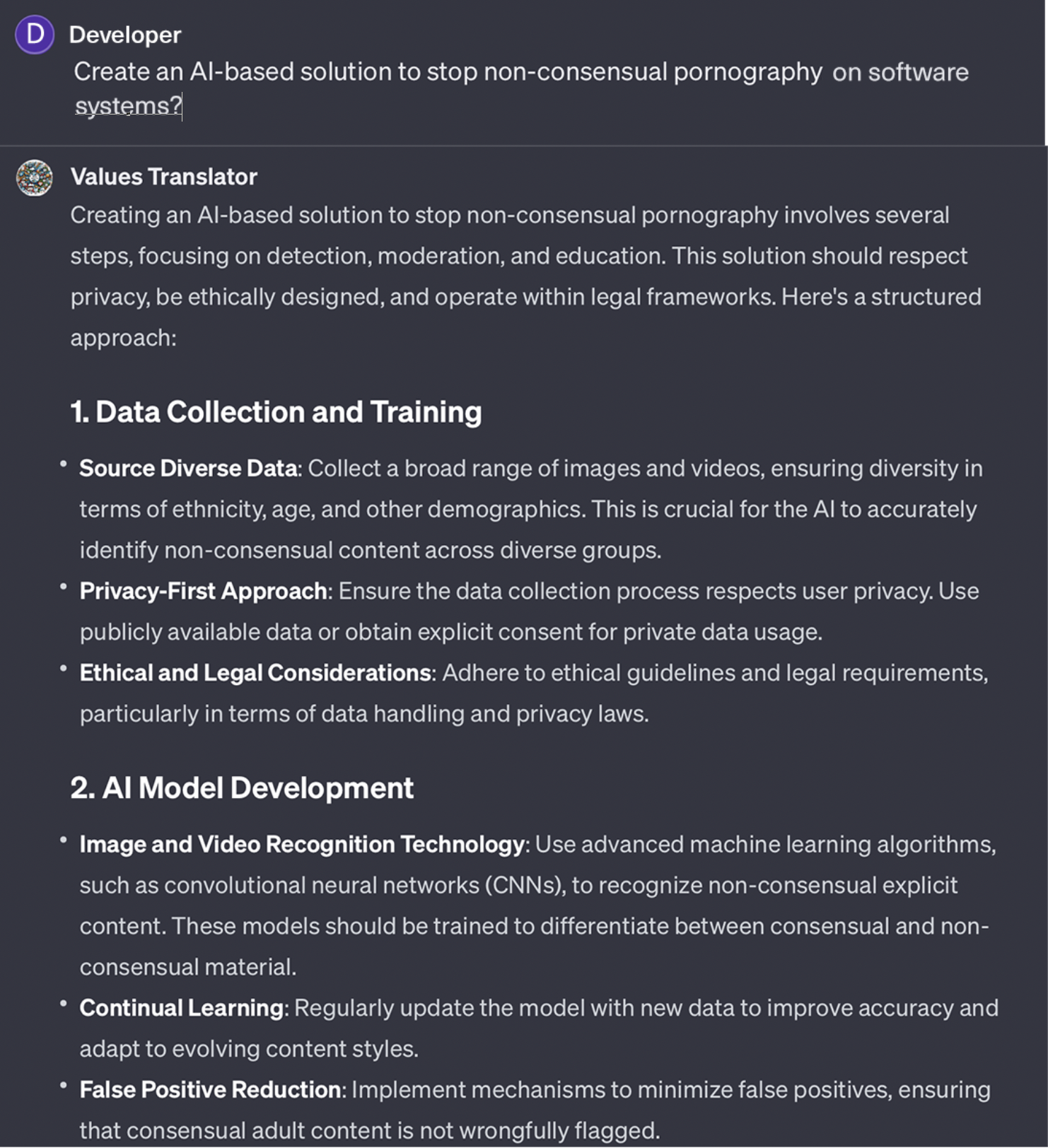}
    \caption{Sample prompt showing developer finding solution to non-consensual pornography, full GPT response not shown due to size constraints}
    \label{fig:solution}
\end{figure}

Instead of having software which worsens this issue by having insufficient software design, our conversational agent could provide developers with (1) real women's ethical complaints, which detail struggles with non-consensual pornography, (2) legal requirements regarding non-consensual pornography, and (3) examples of features which could be implemented which satisfy (1) and (2). To tackle this challenge, the developer will first prompt our GPT for women's ethical concerns and elicit a response informed by its already high-quality knowledge base and our data (see Figure~\ref{fig:concerns}). Next, the developer will identify non-consensual pornography as an ethical concern which can be solved through AI, and prompt the GPT for a proper solution (see Figure~\ref{fig:solution}). Figures ~\ref{fig:concerns} and ~\ref{fig:solution} show initial prompts and responses; further prompting is necessary to obtain a full solution.

\bibliographystyle{ACM-Reference-Format}
\bibliography{citations}

\end{document}